\pdfoutput=1
\documentclass[ejs]{imsart}

\RequirePackage[numbers]{natbib}
\RequirePackage[colorlinks,citecolor=blue,urlcolor=blue]{hyperref}

\doi{10.1214/154957804100000000}
\pubyear{0000}
\volume{0}
\firstpage{0}
\lastpage{0}

\startlocaldefs
\endlocaldefs

\begin{document}

\begin{frontmatter}

\title{A Generalized Savage-Dickey Ratio\thanksref{t1}}
\thankstext{t1}{This is an original survey paper}
\runtitle{A Generalized SDR}

\author{\fnms{Ewan} \snm{Cameron}\corref{}\ead[label=e1]{dr.ewan.cameron@gmail.com}\ead[label=e2,url]{astrostatistics.wordpress.com}}
\address{\printead{e1}\\ \printead{e2}}

\runauthor{Cameron}

\begin{abstract}
In this brief research note I present a generalized version of the Savage-Dickey Density Ratio for representation of the Bayes factor (or marginal likelihood ratio) of nested statistical models; the new version takes the form of a Radon-Nikodym derivative and is thus applicable to a wider family of probability spaces than the original (restricted to those admitting an ordinary Lebesgue density).  A derivation is given following the measure-theoretic construction of Marin \& Robert (2010), and the equivalent estimator is demonstrated in application to a distributional modeling problem.
\end{abstract}




\end{frontmatter}

\section{Introduction}
The Savage-Dickey Density Ratio (SDDR; Dickey 1971 \cite{dic71}) is known as a special formulation of the Bayes factor applicable to the case of nested models.  By far its greatest popularity is in applications to problems in physics and astronomy, such as cosmological model selection (e.g.\ Trotta 2007 \cite{tro07};  Verde et al.\ 2013 \cite{ver13}).  In its original version it is supposed that there exist two competing models coupled to the observed data, $x$, via likelihood function, $f(x|\theta,\psi)$, with the simpler model specified by a fixed $\theta_0$ and prior probability density, $\pi_0(\psi)$, on the nuisance parameter, $\psi$, while the more complex model allows for varying $\theta$ with a joint prior density, $\pi_1(\theta,\psi)$.  According to the SDDR the Bayes factor comparing these models, i.e.,
\[B_{01} = \frac{\int f(x|\theta_0,\psi)\pi_0(\psi)d\psi}{\int f(x|\theta,\psi)\pi_1(\theta,\psi)d\theta d\psi} = \frac{m^{(0)}(x)}{m^{(1)}(x)}\]
(with domain of integration the natural domain of $\psi$ or $\theta\times\psi$, as appropriate), may be rewritten without direct reference to the simpler model as 
\begin{equation}\label{sddr}B_{01} = \frac{\pi_1(\theta_0|x)}{\pi_1(\theta_0)} \mathrm{\ provided\ that\ } \pi_1(\psi|\theta_0) = \pi_0(\psi).\end{equation}

However, as Marin \& Robert (2010) \cite{mar10} have pointed out, Equation \ref{sddr} in fact presents a mathematically void constraint since the conditional probability is defined only up to sets of measure zero: with $\{\theta_0\}$ being one such set the definition of $\pi_1(\psi|\theta_0)$ is effectively arbitrary, i.e., can be imposed regardless of $\pi_1(\psi|\theta)$.  Accordingly, these authors demonstrate that the SDDR construction is effectively tautological from a measure-theoretic perspective owing to the availability of specific versions of the posterior that make it hold (or otherwise); while the stronger constraint of a separable $\pi_1(\theta,\psi)=\pi_1(\theta)\pi_0(\psi)$ partially alleviates the ambiguity and suggests a novel estimator for $B_{01}(x)$ under non-separable $\pi_1(\theta,\psi)$ given the availability of simulations from a separable auxiliary.

With various problems of interest for Bayesian statistics featuring priors on (perhaps infinite-dimensional) probability spaces not amenable to representation as probability densities with respect to the Lebesgue reference measure (e.g. for  model selection under Dirichlet process or Gaussian process priors) one may naturally enquire whether an equivalent to the SDDR exists for such cases?  Following the notation and presentation of Marin \& Robert (2010) \cite{mar10} I confirm below in Section \ref{sdr} that a generalized version can indeed be constructed in the tautological sense, though it no longer takes the form of a density ratio, becoming instead a Radon-Nikodym derivative.  Moreover, an equivalent to their novel SDDR-based estimator given a separable auxiliary can also be derived and, as I demonstrate in Section \ref{numerical} through a numerical example, can be readily applied for practical Bayes factor estimation.

\section{A Generalized Savage-Dickey Ratio}\label{sdr}
A generalized version of the SDDR (which, as motivated above, is no longer a \textit{density} ratio; hence, SDR) may be derived thus in three steps:
\[B_{01}(x) = \frac{\int f(x|\theta_0,\psi) \{dP_\psi^{(0)}(\psi)\}}{\int f(x|\theta,\psi)\{dP^{(1)}_{\theta,\psi}(\theta,\psi)\}} \mathrm{\ \ [by\ definition]}\]
\[= \frac{\int f(x|\theta_0,\psi) \{dP_{\psi|\theta_0}^{(1)}(\psi)\}}{m^{(1)}(x)} \mathrm{\ \ [using\ a\ specific\ version\ of\ }P_{\psi|\theta}^{(1)}]\]
\[= \frac{dP^{(1)}_{\theta|x}}{dP^{(1)}_{\theta}}(\theta_0)\mathrm{\ \ [using\ a\ specific\ version\ of\ }\frac{dP^{(1)}_{\theta|x}}{dP^{(1)}_{\theta}}].\]
The integration denoted here by $\int f \{dP\}$ is, of course, Lebesgue integration with $f$ assumed to be a real, measurable function.  The specific version of the conditional probability on the second line is evidently a $P_{\psi|\theta}^{(1)}$ defined such that \[\mathrm{P}^{(1)}(A_1 \times A_2) = \int_{A_1 \times A_2} \{dP^{(1)}_{\theta,\psi}(\theta,\psi)\} = \int_{A1} \left[ \int_{A2} \{dP^{(1)}_{\psi|\theta}(\psi)\} \right](\theta) \{dP^{(1)}_\theta(\theta)\} \]  for all $A_1 \times A_2$ in the $\sigma$-algebra of the $\theta\times\psi$ space with the particular choice of $P_{\psi|\theta_0}^{(1)}(\psi) = P^{(0)}_\psi(\psi)$ on the set of zero $P^{(1)}_{\theta}$-measure, $\{\theta_0\}$.  Likewise, the specific version of the Radon-Nikodym derivative on the second line required to ensure the stated equality is a $\frac{dP^{(1)}_{\theta|x}}{dP^{(1)}_{\theta}}$ defined such that \[ \mathrm{P}^{(1)}_{\theta|x}(A_1) = \int_{A_1} \{dP^{(1)}_{\theta|x}(\theta)\} = \int_{A_1} \frac{dP^{(1)}_{\theta|x}}{dP^{(1)}_\theta}(\theta) \{dP^{(1)}_\theta(\theta)\} \]
for all $A_1$ in the $\sigma$-algebra of the $\theta$ space with the particular choice of $\frac{dP^{(1)}_{\theta|x}}{dP^{(1)}_{\theta}}(\theta_0) = \int \frac{f(x|\theta_0,\psi)}{m^{(1)}(x)} \{dP^{(1)}_{\psi|\theta_0}(\psi)\}$  on the same.  Appropriate measure-theoretic definitions of the conditional probability and Radon-Nikodym derivative (as used above) are given by, e.g., Halmos \& Savage (1950) \cite{hal50}.  With the Radon-Nikodym derivative (a.e.) equal to the ratio of densities for probability spaces admitting a Lebesgue reference measure, the equivalence between the generalized SDR and the original SDDR is immediate, $\frac{dP_{\theta|x}^{(1)}}{dP_\theta^{(1)}}(\theta_0) = \frac{\pi_1(\theta_0|x)}{\pi_1(\theta_0)}$ (again, for the implied specific choices of each).

As shown by Marin \& Robert (2010) \cite{mar10}, the tautological nature of the SDDR is somewhat alleviated by specifying a separable $\pi_1(\theta,\psi)=\pi_1(\theta)\pi_0(\psi)$; leading to their novel estimator for arbitrary $\pi_1(\theta,\psi)$ given a separable auxiliary, $\tilde{\pi}_1(\theta,\psi)$.  Likewise in the case of the generalized SDR, for which an equivalent estimator  may be recovered via much the same arguments, supposing first the separable auxiliary, $\tilde{P}^{(1)}_{\theta,\psi}$, taking the form of a product measure, $\{d\tilde{P}^{(1)}_{\theta,\psi} (\theta,\psi)\} = \{dP^{(1)}_{\theta}(\theta)\}\times\{dP^{(0)}_{\psi}(\psi)\}$.  With $\tilde{m}^{(1)}(x)$ its corresponding marginal likelihood, the Bayes factor becomes
\[
B_{01}(x) = \left[\frac{d\tilde{P}^{(1)}_{\theta|x}}{dP^{(1)}_{\theta}}(\theta_0)\right] \frac{\tilde{m}^{(1)}(x)}{m^{(1)}(x)}.
\]
Now an unbiased estimator for the left term given a sample of size $N$ from $\tilde{P}^{(1)}_{\theta,\psi|x}$ is simply 
\[
\frac{1}{N} \sum_{i=1}^N \frac{d\tilde{P}^{(1)}_{\theta|\psi_i,x}}{dP^{(1)}_{\theta}}(\theta_0),
\]
and an unbiased estimator for the right term given a sample of size $N$ from $P^{(1)}_{\theta,\psi|x}$ is likewise
\[\frac{1}{N}\sum_{i=1}^N\frac{dP_{\psi}^{(0)}}{dP^{(1)}_{\psi|\theta_i}}(\psi_i),
\]
such that their product yields an unbiased estimator for $B_{01}$ itself.  I illustrate the application of this estimator in the numerical example presented below.

\section{Numerical Example: Non-parametric Density Estimation for the Old Faithful Dataset}\label{numerical}
An interesting case study for model selection under non-Lebesgue density priors is provided by the problem of non-parametric density estimation via the Dirichlet process (e.g.\ Escobar \& West 1995 \cite{esc95}; Doss 2012 \cite{dos12}).  As a somewhat artificial example, suppose for data, $x$, the following hierarchical structure defining the more complex of our candidate generative models:
\begin{eqnarray}\nonumber
x_i\ (i=1,\ldots,n) &\sim& f_{\mathcal{N}}(\cdot|\{\mu_i\},k/\alpha)\\
\{\mu_i\}\ (i=1,\ldots,n) &\sim& P\nonumber\\
P &\sim& \mathrm{DP}(\alpha,\mathcal{N}(m,\Sigma))\nonumber\\
\alpha &\sim& \Gamma(\nu_1,\nu_2), \nonumber
\end{eqnarray}
where DP represents the Dirichlet process with concentration index, $\alpha$, and Normal reference density, $\mathcal{N}(m,\Sigma)$; the latter characterized by its mean, $m$, and variance, $\Sigma$.  Accordingly, $P$ represents a single realization from the DP thus specified and $f_\mathcal{N}$ represents the density of the Normal (kernel) likelihood function given the list of (component) means, $\{\mu_i\}$, and the shared variance term, $k/\alpha$ (here $k$ is fixed).  Integration over the second and third layers of this hierarchical set up yields a prior, $P^{(1)}_{\alpha,\{\mu_i\}}$, that is evidently \textit{not} absolutely continuous with respect to Lebesgue measure; as Doss (2012) \cite{dos12} notes, the DP prior assigns a non-zero probability to the event that some of the $\{\mu_i\}$ are exactly equal.  Moreover, this prior cannot be decomposed as a simple product measure on the space of $\alpha \times \{\mu_i\}$ owing to the dependence structure of $\alpha$ appearing as both the concentration index of the DP and as a precision factor in the shared variance term.

Suppose further that under our simpler alternative model we have $\alpha$ fixed to some $\alpha_0$, yielding again a non-Lebesgue density prior, $P^{(0)}_{\{\mu_i\}}$, on the `nuisance parameter', $\{\mu_i\}$.  To apply the generalized SDR version of the Marin \& Robert (2010) \cite{mar10} estimator to this problem requires the introduction of a product measure proxy, $\tilde{P}^{(1)}_{\alpha,\{\mu_i\}}$, defined such that $d\tilde{P}^{(1)}_{\alpha,\{\mu_i\}}(\alpha,\{\mu_i\}) = dP^{(1)}_{\alpha}(\alpha) \times dP^{(0)}_{\{\mu_i\}}(\{\mu_i\})$, which may be constructed simply by decoupling the $\alpha$ from the DP in the above hierarchical definition of the more complex model, replacing $\mathrm{DP}(\alpha,\mathcal{N}(m,\Sigma))$ with $\mathrm{DP}(\alpha^\prime,\mathcal{N}(m,\Sigma))$ and adding the extra layer, $\alpha^\prime \sim \Gamma(\nu_1,\nu_2)$.  Simulation from $\tilde{P}^{(1)}_{\alpha,\{\mu_i\}|x}$ is easily achieved via Gibbs sampling using existing formulae for the conjugate conditionals of the DP random effects model (cf.\ Burr \& Doss 2005 \cite{bur05}) and for the Normal with known mean but unknown variance; likewise for simulation from $P^{(1)}_{\alpha,\{\mu_i\}|x}$.  

Here I adopt as benchmark dataset the record of 272 Old Faithful eruption times available in \texttt{R} as the first column of \texttt{data(faithful)}.  A histogram inspection of the data suggests the presence of at least two separate modes, one centered on $\sim$2 and another on $\sim$4.5.  For the hyperparameters controlling $P^{(1)}_{\alpha,\{\mu_i\}}$ and $P^{(0)}_{\{\mu_i\}}$ I take $\{m=3,\Sigma=4,\nu_1 = 5,\nu_2=5,k=1\}$, and for the simpler model I have chosen the fixed point of $\alpha_0 = 0.5$.  Thirty repeat simulations to compute the generalized SDR version of Marin \& Robert's (2010) \cite{mar10} Bayes factor estimator with $N=5,000$ draws from each of $\tilde{P}_{\alpha,\{\mu_i\}|x}^{(1)}$ and $P_{\alpha,\{\mu_i\}|x}^{(1)}$ yields a mean of $\hat{B}_{01} = 3.9$ with a standard deviation of 0.4, indicating a mild preference for the simpler model at $\alpha_0=0.5$.  

It is worth noting, however, that a number of other well-known marginal likelihood estimation techniques are also readily amenable to Bayes factor estimation under stochastic process priors, including Chib's method (cf.\ Chib 1995 \cite{chi95}; Basu \& Chib 2003 \cite{bas03}) and biased sampling (cf.\ Vardi 1985 \cite{var85}; Doss 2012 \cite{dos12}; Cameron \& Pettitt 2013a,b \cite{cam13a,cam13b})---and for more complex problems it is to be expected that strategies, such as the latter, that facilitate the estimation via a sequence of flexible bridging distributions (e.g.\ the thermodynamic integration pathway or the partial data posteriors pathway) will ultimately prove more computationally efficient.  Nevertheless, the challenges of marginal likelihood estimation under stochastic process priors are many and varied, and the generalized SDR may yet prove to be a feasible, practical option for small scale problems such as that of the numerical example given here.

\section{Conclusions}
There exists a generalized version of the SDDR for probability measures over spaces not anemable to representation as probability densities with respect to the Lebesgue measure; the generalized version being available instead as a particular (tautological) choice of the Radon-Nikodym derivative between posterior and prior for the parameter of interest under the more complex of two nested models (marginalized over any nuisance parameters).  An equivalent version of Marin \& Robert's (2010) \cite{mar10} SDDR-based estimator for the posterior Bayes factor via a separable auxiliary is also available, and has been demonstrated in application here to a numerical example concerning nonparametric density estimation for the Old Faithful dataset.

\end{document}